\documentclass[12pt,a4paper]{article}
\usepackage{amsmath}
\usepackage[utf8]{inputenc}
\usepackage[hidelinks]{hyperref}
\usepackage{german}
\usepackage{color}
\usepackage{nicefrac}
\usepackage{siunitx}
\usepackage{graphicx}
\usepackage{listings}
\usepackage{xcolor}
\newcommand*{\ubstrut}{\vphantom{\vrule depth 1.5mm}}
\graphicspath{{./pics/}}
\medskipamount  
\definecolor{code_color}{RGB}{230,230,230}
\newcommand{\MyListing}[1]%
  {\begin{lstlisting}[keepspaces=true,basicstyle=\footnotesize,%
                      backgroundcolor=\color{code_color},language=#1]}

\newlength{\overwritelength}
\newlength{\minimumoverwritelength}
\newcommand{\overwrite}[3][brown]{%
  \settowidth{\overwritelength}{$#2$}%
  \ifdim\overwritelength<\minimumoverwritelength%
    \setlength{\overwritelength}{\minimumoverwritelength}\fi%
  \stackrel
    {%
      \begin{minipage}{\overwritelength}%
        \color{#1}\centering\small #3\\%
        \rule{1pt}{9pt}%
      \end{minipage}}
    {\colorbox{#1!50}{\color{black}$\displaystyle#2$}}}

\begin{document}
\begin{center}
\Large      Simulation von Wellenmaschinen mit                  \\
            GNU Octave, Python und C++                          \par
\normalsize Tilman Küpper, tilman.kuepper@hm.edu                \\
            Hochschule für angewandte Wissenschaften München    \par
            17. Oktober 2017
\end{center}

% -----------------------------------------------------------------------------
% Zusammenfassung
% -----------------------------------------------------------------------------
\begin{abstract}
In "`Die Wellenmaschine -- Grundlagen der Wellenausbreitung, Dispersion,
Reflexion, Simulation"' \cite{kuepper2015} werden die in Schule und
Hochschule verbreiteten Wellenmaschinen näher betrachtet und die
Wellenausbreitung auf solchen Maschinen mithilfe von MATLAB simuliert.
Die kommerzielle Software MATLAB ist freilich nicht überall verfügbar,
sodass sich die Frage nach freien Alternativen stellt.

Der MATLAB-Quelltext im genannten Artikel kann zum Beispiel
mit GNU Octave ausgeführt werden. Die im Vergleich zu MATLAB geringere
Ausführungsgeschwindigkeit lässt sich durch Umformung des zu lösenden
Differentialgleichungssystems in Matrixschreibweise zumindest teilweise
kompensieren.

Aber auch mit klassischen Programmiersprachen wie Python oder C++
lassen sich diese und ähnliche Simulationen leicht durchführen.
Passende Bibliotheken zur Arbeit mit Vektoren und Matrizen, zum
Lösen von Differentialgleichungen und zur animierten Darstellung
der Simulationsergebnisse werden im Artikel vorgestellt.
\end{abstract}

% -----------------------------------------------------------------------------
% 1. Wellenmaschinenmodell
% -----------------------------------------------------------------------------
\section{Wellenmaschinenmodell}
Ausgangspunkt für die folgenden Betrachtungen sind das in \cite{kuepper2015}
vorgestellte vereinfachte Modell einer Wellenmaschine (Abbildung \ref{modell})
aus $N$ Kugeln der Masse $m$, die mit Federn verbunden sind sowie die daraus
abgeleiteten Bewegungsgleichungen:%
\footnote{Siehe \cite{kuepper2015}, Abbildung 2 und Gleichung 1}
%
\begin{equation}
\label{bewegungsgleichungen}
\begin{array}{r@{\ }l}
    m \ddot{y}_n &= F_\mathrm{T} = F_\mathrm{L} + F_\mathrm{R}  \\[3mm]
    m \ddot{y}_n &= D(y_{n-1} - y_n) + D(y_{n+1} - y_n)         \\[3mm]
    \ddot{y}_n   &= \dfrac{D}{m} ( y_{n-1} + y_{n+1} - 2y_n )
\end{array}
\end{equation}
\pagebreak

Die Kugel am linken Rand der Wellenmaschine ($n = 1$) wird von außen 
zum Beispiel sinusförmig angeregt, auf die Kugel am rechten Rand
($n = N$) wirkt geschwindigkeitsabhängige Reibung (Reibungskoeffizient
$\alpha$):%
\footnote{Bei der Vorbereitung dieses Artikels ist aufgefallen,
dass sich in \cite{kuepper2015}, Formel 29, innerhalb der Klammer ein
Vorzeichenfehler eingeschlichen hat. Der in Anhang A abgedruckte
Quelltext ist allerdings korrekt.}
%
\begin{equation}
\begin{array}{r@{\ }l}
  \ddot{y}_1 &= \dfrac{D}{m} ( \hat{Y} \sin(\omega t) - 2 y_1 + y_2 )        \\[4mm]
  \ddot{y}_N &= \dfrac{D}{m} ( y_{N-1} - y_N ) - \dfrac{\alpha}{m} \dot{y}_N  \\[4mm]
\end{array}
\end{equation}

\begin{figure}[t] 
  \centering 
  \includegraphics[scale=0.4]{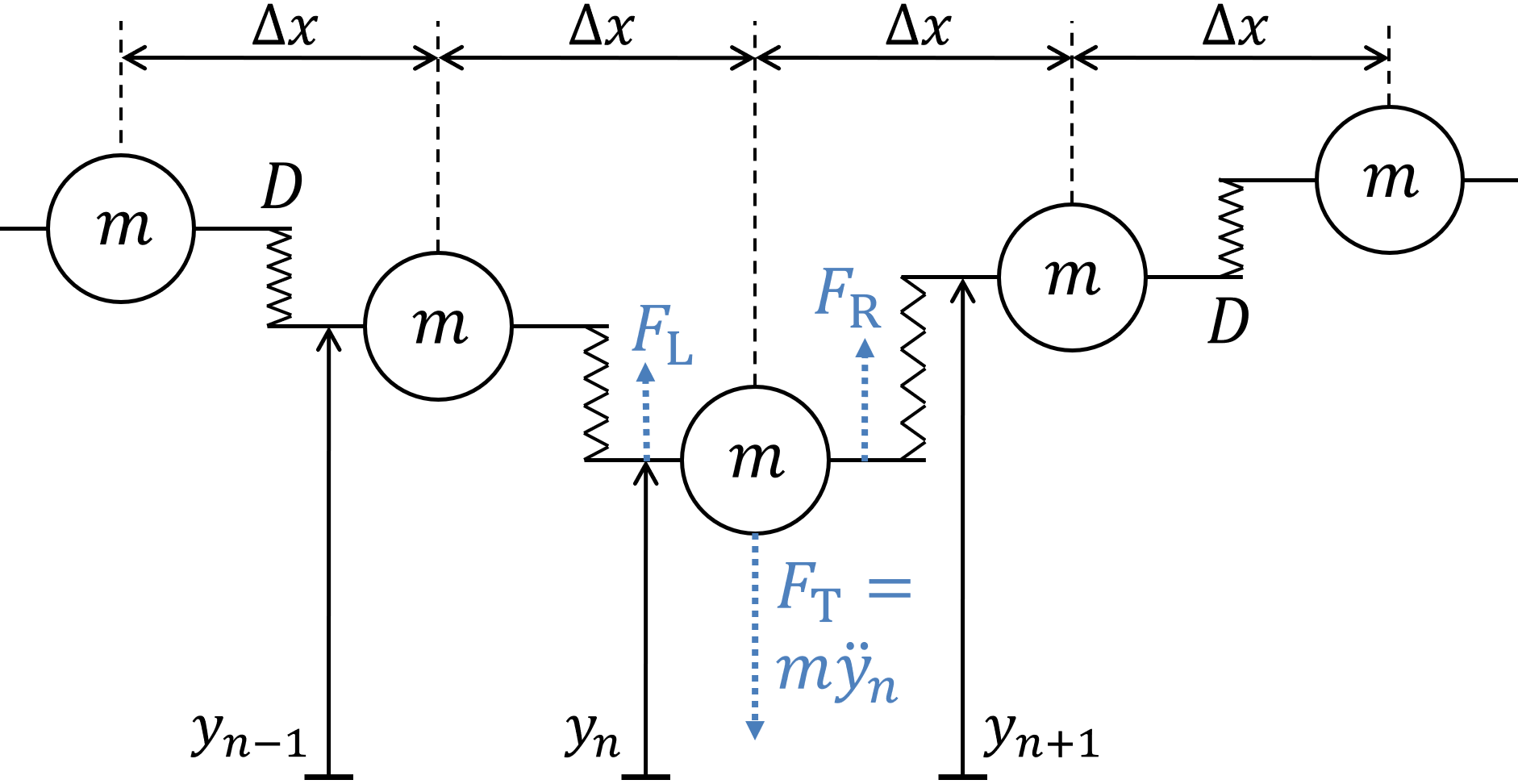}
  \caption{Vereinfachtes Wellenmaschinenmodell}
	\label{modell}
\end{figure} 

Insgesamt wird die Wellenmaschine durch ein System von $N$ gewöhnlichen
Differentialgleichungen zweiter Ordnung beschrieben. Zur Simulation
durch einen Computer werden diese in ein System von $2N$
Differentialgleichungen erster Ordnung umgeformt.

Dazu werden die Zustandsvariablen $u_i$ eingeführt. Im Vergleich
zur Darstellung in \cite{kuepper2015} wurde die Reihenfolge der
Zustandsvariablen $u_i$ verändert, um in Gleichung \ref{matrixform}
zu einer möglichst übersichtlichen Matrix $M$ zu gelangen:
%
\begin{equation}
\begin{array}{r@{\ }l}
 \vec{u}              &= (u_1, u_2\ \dots\ u_{2N}) ^ \top \quad  \\[2mm]
 \mbox{mit }\ u_1     &= y_1,\ u_2 = y_2\ \dots\ u_N = y_N \quad \\[2mm]
 \mbox{und }\ u_{N+1} &= \dot{y}_1,\ u_{N+2} = \dot{y}_2\ \dots\ u_{2N} = \dot{y}_N
\end{array}
\end{equation}

Das Gleichungssystem erster Ordnung kann nun angegeben werden:
%
\begin{alignat}{1}
  \parbox[c]{4cm}{\footnotesize\flushright erste Kugel\\am linken Rand\\der Wellenmaschine\\\hfill} &\left\{
  \begin{array}{r@{\ }l}
    \dot{u}_1     &= u_{N+1}  \\[4mm]
    \dot{u}_{N+1} &= -\dfrac{2D}{m} u_1 + \dfrac{D}{m} u_2 + \dfrac{D}{m} \hat{Y} \sin(\omega t)
    \ \ \ \ \ \
  \end{array}
  \right.  \\[4mm]
  %
  \mbox{\footnotesize mittlere Kugeln} &\left\{
  \begin{array}{r@{\ }l}
    \dot{u}_n     &= u_{N+n}  \\[4mm]
    \dot{u}_{N+n} &= \dfrac{D}{m} u_{n-1} - \dfrac{2D}{m} u_{n} + \dfrac{D}{m} u_{n+1}
  \end{array}
  \right.  \\[4mm]
  %
  \parbox[c]{4cm}{\footnotesize\flushright letzte Kugel\\am rechten Rand\\der Wellenmaschine\\\hfill} &\left\{
  \begin{array}{r@{\ }l}
    \ \ \
    \dot{u}_N    &= u_{2N}  \\[4mm]
    \dot{u}_{2N} &= \dfrac{D}{m} u_{N-1} - \dfrac{D}{m} u_N - \dfrac{\alpha}{m} u_{2N}
  \end{array}
  \right.
\end{alignat}

Dasselbe Differentialgleichungssystem in Matrixschreibweise:
%
\begin{equation}
  \label{matrixform}
  \overwrite{\ubstrut\dot{\vec{u}} = M \cdot \vec{u}}%
  {\footnotesize Kugeln und Federn der Wellenmaschine}
  +
  \overwrite{\ubstrut\hat{Y} \sin(\omega t) \cdot \vec{b}}%
  {\footnotesize äußere Anregung}
\end{equation}

Für die $\medmuskip=0mu 2N\times2N$-Matrix $M$ gilt
(nicht abgebildete Elemente sind null):
%
\begin{equation}
\begin{split}
  & M = \\ 
  & \left( \small
  \begin{array}{rrrrrrr|ccccc}
    & & & & & & &                       1 &         &   \\
    & & & & & & &                         &  \ddots &   \\
    & & & & & & &                         &         & 1 \\
    \hline
    \nicefrac{-2D}{m} & \nicefrac{  D}{m} &                   &                 &                   &                   &                  &        & &                    \\
    \nicefrac{  D}{m} & \nicefrac{-2D}{m} & \nicefrac{D}{m}   &                 &                   &                   &                  &        & &                    \\
                      & \nicefrac{  D}{m} & \nicefrac{-2D}{m} & \nicefrac{D}{m} &                   &                   &                  &        & &                    \\
                      &                   &                   & \ddots          &                   &                   &                  &        & &                    \\
                      &                   &                   & \nicefrac{D}{m} & \nicefrac{-2D}{m} & \nicefrac{  D}{m} &                  &        & &                    \\
                      &                   &                   &                 & \nicefrac{  D}{m} & \nicefrac{-2D}{m} & \nicefrac{ D}{m} &        & &                    \\
                      &                   &                   &                 &                   & \nicefrac{  D}{m} & \nicefrac{-D}{m} &        & & \nicefrac{-\alpha}{m}
  \end{array}
  \right)
\end{split}
\end{equation}

Beim Vektor $\vec{b}$ ist lediglich das
$\medmuskip=0mu (N+1)$-te Element ungleich null:
%
\begin{equation}
  \vec{b} = \left(
  \begin{array}{rrr|cccc}
    0 & \dots & 0 &         \nicefrac{D}{m} & 0 & \dots & 0
  \end{array}
  \right) ^ \top
\end{equation}

% -----------------------------------------------------------------------------
% 2. Simulation mit GNU Octave
% -----------------------------------------------------------------------------
\section{Simulation mit GNU Octave}

Das Differentialgleichungssystem in Matrixschreibweise (Gleichung
\ref{matrixform}) kann leicht in entsprechenden Quelltext für MATLAB
bzw. GNU Octave umgesetzt werden. Insbesondere bei der Verwendung von
GNU Octave, einer freien MATLAB-Alternative \cite{octave}, führt
die Matrixschreibweise zu einer deutlich schnelleren Verarbeitung
als die in \cite{kuepper2015}, Anhang A, gezeigte Variante.%
\footnote{Rechenzeit auf einem HP Spectre 13 mit i7-4500U-Prozessor
und \SI{8}{GB} RAM: \SI{5}{s} mit Matrixschreibweise im Vergleich zu
\SI{35}{s} für die mit ode45 gerechnete Variante aus \cite{kuepper2015}.}

\MyListing{MATLAB}
% Simulation einer Wellenmaschine mit MATLAB
function wellensim()
    global M b FREQ AMPL

    N     = 20;    % Kugelanzahl
    MASSE = 0.01;  % Kugelmasse
    FEDER = 1.0;   % Federkonstante
    FREQ  = 0.5;   % Anregungsfrequenz
    AMPL  = 1.0;   % Anregungsamplitude
    REIB  = sqrt(FEDER * MASSE);  % Reibung
    
    % Matrix M belegen
    M = zeros(2 * N, 2 * N);
    for i = 1:N
        M(i, N + i) = 1;
        M(N + i, i) = -2 * FEDER / MASSE;
        if i > 1
            M(N + i, i - 1) = FEDER / MASSE;
        end
        if i < N
            M(N + i, i + 1) = FEDER / MASSE;
        end
    end
    M(2 * N, N) = -FEDER / MASSE;
    M(2 * N, 2 * N) = -REIB / MASSE;
    
    % Vektor b belegen
    b = zeros(2 * N, 1);
    b(N + 1) = FEDER / MASSE;

    % DGL-System integrieren
    t_interval = 0:0.05:100;
    anf = zeros(1, 2 * N);
    [T, Y] = ode45(@dgl_system, t_interval, anf);
\end{lstlisting}
\pagebreak    
\MyListing{MATLAB}
    for i = 1:length(T)
        plot(Y(i, 1:N), 'bo-');
        ylim([-1.5 1.5]);
        title(sprintf('t = 
        grid on;
        drawnow;
        pause(0.1);
    end
end

function dudt = dgl_system(t, u)
    global M b FREQ AMPL
    omega = 2 * pi * FREQ;
    dudt = M * u + b * AMPL * sin(omega * t);
end
\end{lstlisting}

\begin{figure}[t] 
  \centering 
  \includegraphics[width=1.0\linewidth]{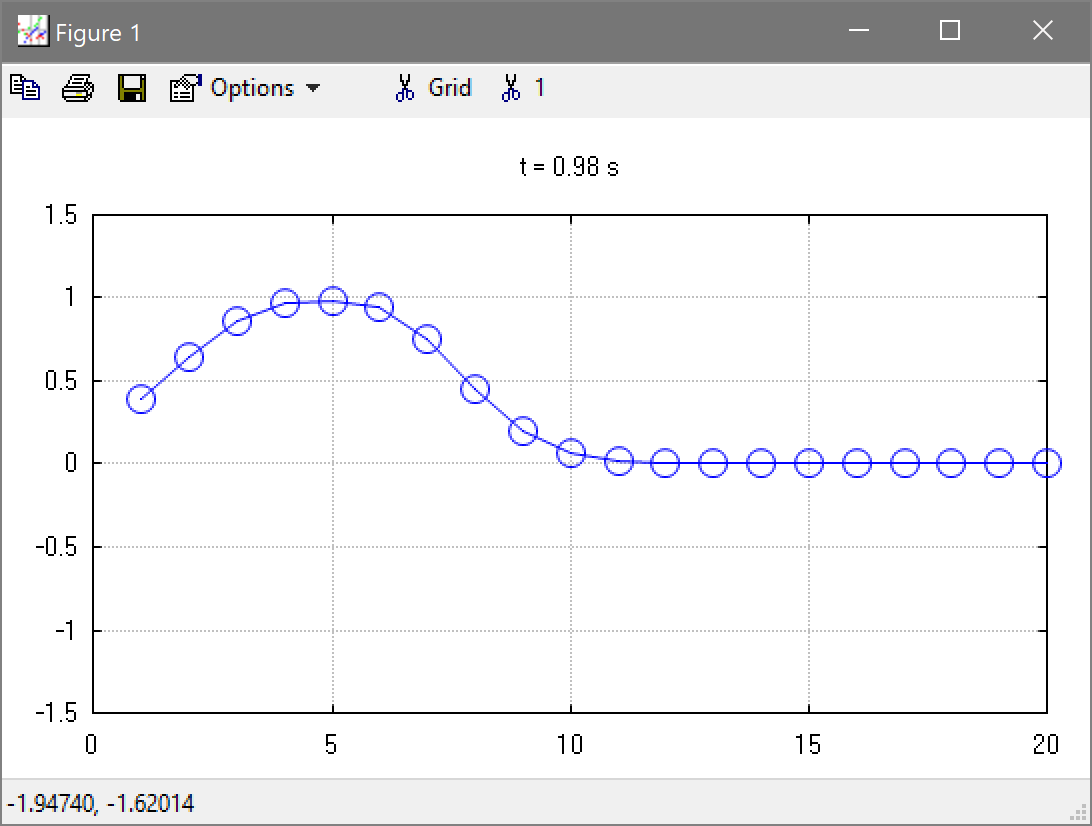}
  \caption{Simulation mit GNU Octave}
\end{figure} 

\section{Simulation mit Python}

Python ist wie MATLAB/Octave eine interpretierte Programmiersprache
mit dynamischem Typsystem und automatischer Speicherverwaltung.
Der kompakte Sprachkern kann durch unzählige Bibliotheken
erweitert werden, die ebenso wie Python selbst oft kostenlos
und mit vollständigen Quellen im Internet verfügbar sind. Für
technisch-wissenschaftliche Anwendungen sind die Bibliotheken
SciPy, NumPy und Matplotlib\footnote{Siehe 
\href{https://www.scipy.org}{https://www.scipy.org},
\href{http://www.numpy.org}{http://www.numpy.org} und
\href{http://matplotlib.org}{http://matplotlib.org}}
besonders interessant: Sie stellen Funktionalitäten zur Arbeit mit Matrizen
und Vektoren, zum Lösen von Differentialgleichungen und für die grafische
Ergebnisdarstellung zur Verfügung, vergleichbar mit denen von MATLAB/Octave.
\cite{python}

Wie schon bei der Simulation mittels MATLAB/Octave wird das
Differentialgleichungssystem der Wellenmaschine in einer Funktion
(dgl\_system) definiert. Die numerische Integration dieses Systems erfolgt
durch Aufruf der SciPy-Funktion integrate.odeint, die animierte Darstellung
der Simulationsergebnisse erfolgt mittels Matplotlib. Der größte Unterschied
im Vergleich zur Programmierung mit MATLAB/Octave dürfte neben
der etwas unterschiedlichen Syntax der Zugriff auf einzelne Matrix-
und Vektorelemente sein, welcher in Python unter Verwendung eines
nullbasierten Index geschieht.

\MyListing{Python}
# Simulation einer Wellenmaschine in Python
import numpy as np
import matplotlib.pyplot as plt
import scipy.integrate as integrate 

N     = 20;    # Kugelanzahl
MASSE = 0.01;  # Kugelmasse
FEDER = 1.0    # Federkonstante
FREQ  = 0.5    # Anregungsfrequenz
AMPL  = 1.0    # Anregungsamplitude
REIB  = np.sqrt(FEDER * MASSE)  # Reibung

# Matrix M belegen
M = np.zeros((2 * N, 2 * N))
for i in range(0, N):
    M[i, N + i] = 1
    M[N + i, i] = -2 * FEDER / MASSE
    if i > 0:
        M[N + i, i - 1] = FEDER / MASSE
    if i < N - 1:
        M[N + i, i + 1] = FEDER / MASSE
M[2 * N - 1, N - 1] = -FEDER / MASSE
M[2 * N - 1, 2 * N - 1] = -REIB / MASSE
\end{lstlisting}
\pagebreak    
\MyListing{Python}
# Vektor b belegen
b = np.zeros((2 * N))
b[N] = FEDER / MASSE

# DGL-System der Wellenmaschine
def dgl_system(u, t):
    omega = 2 * np.pi * FREQ
    dudt = M @ u + b * AMPL * np.sin(omega * t)
    return dudt

# DGL-System integrieren
t_interval = np.linspace(0, 100, 2001)
anf = np.zeros((2 * N))
Y = integrate.odeint(dgl_system, anf, t_interval)

# Animation der Simulationsergebnisse
for i in range(0, len(t_interval)):
    plt.gcf().clear()
    plt.plot(Y[i, 0:N], "bo:")
    plt.ylim((-1.5, 1.5));
    plt.grid(True)
    plt.title("t = 
    plt.pause(0.1)
\end{lstlisting}

\begin{figure}[t] 
  \centering 
  \includegraphics[width=1.0\linewidth]{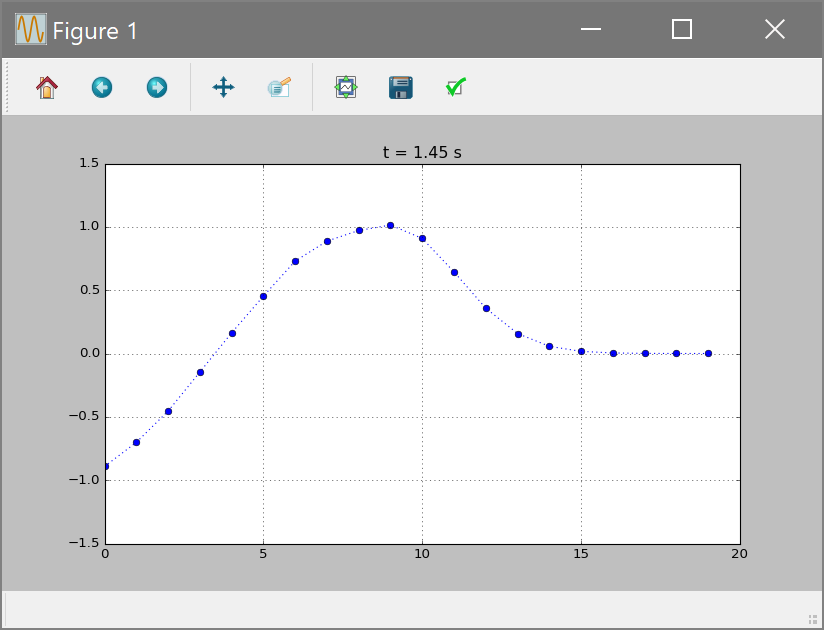}
  \caption{Simulation mit Python}
\end{figure} 

\section{Simulation mit C++}

Auch bei der Programmiersprache C++ gehören nicht alle zur Simulation der
Wellenmaschine (und der grafischen Ergebnisdarstellung) benötigten Klassen
und Funktionen zum standardisierten Sprachkern. Wie aber bei der Programmierung
mit Python sind auch für C++ unzählige geeignete Bibliotheken frei im Internet
verfügbar.

Für Anwendungen aus dem Bereich der linearen Algebra sind dies etwa
die Bibliotheken Boost uBLAS, Armadillo und Newmat\footnote{Siehe
\href{http://www.boost.org/doc/libs/1\_65\_0/libs/numeric/ublas/doc/index.html}%
     {http://www.boost.org/doc/libs/1\_63\_0/libs/numeric/ublas/doc/index.html},\\
\href{http://arma.sourceforge.net}{http://arma.sourceforge.net} und 
\href{http://www.robertnz.net/nm\_intro.htm}%
     {http://www.robertnz.net/nm\_intro.htm}},
die numerische Lösung von Differentialgleichungssystemen ist mit
Boost odeint oder der GNU Scientific Library möglich\footnote{Siehe
\href{https://www.gnu.org/software/gsl}{https://www.gnu.org/software/gsl} und\\
\href{http://www.boost.org/doc/libs/1\_65\_0/libs/numeric/odeint/doc/html/index.html}%
     {http://www.boost.org/doc/libs/1\_65\_0/libs/numeric/odeint/doc/html/index.html}},
zur grafischen Ergebnisdarstellung können Chart-Module wie Qt Charts oder
TeeChart genutzt werden oder auch Schnittstellen zu Gnuplot\footnote{Siehe
\href{https://doc.qt.io/qt-5/qtcharts-index.html}%
     {https://doc.qt.io/qt-5/qtcharts-index.html},
\href{https://www.steema.com}{https://www.steema.com} und\\
\href{https://code.google.com/archive/p/gnuplot-cpp}%
     {https://code.google.com/archive/p/gnuplot-cpp}}.

Das folgende Programmbeispiel basiert auf den Bibliotheken HMMatrix,
HMDGL und HMChart, die auf der Homepage des Autors\footnote{Siehe
\href{http://kuepper.userweb.mwn.de/software.htm}%
     {http://kuepper.userweb.mwn.de/software.htm}}
zum Download bereitstehen. Zur Programmierung mit diesen Bibliotheken ist
keine spezielle Installation erforderlich, nur einige wenige Dateien
müssen zum C++-Projekt hinzugefügt werden. Diese Bibliotheken sind daher
auch für C++-Einsteiger gut geeignet, sie können zudem mit Microsoft
Windows, Mac OS X wie auch unter GNU/Linux gleichermaßen eingesetzt werden.

Der Programmaufbau ist im Übrigen mit den bereits gezeigten Beispielen in
MATLAB/Octave und Python vergleichbar: Das zu lösende Differentialgleichungs%
system ist als C++-Funktion (sys) definiert, zur numerischen Integration wird
durch Aufruf der Funktion dgl\_rk4 das klassische Runge-Kutta-Verfahren
verwendet, nach jedem Simulations-Zeitschritt (step) werden die aktuellen
Kugelpositionen zur grafischen Darstellung zum Chart-Modul übertragen.
Der Zugriff auf einzelne Matrix- und Vektorelemente geschieht auch in
C++ unter Verwendung eines nullbasierten Index.

\pagebreak
\MyListing{C++}
// Simulation einer Wellenmaschine in C++
#define _USE_MATH_DEFINES

#include <math.h>
#include "chart.h"
#include "hmdgl.hpp"
#include "matrix.hpp"

using namespace std;
using namespace HMDGL;
using namespace HMMatrix;

const auto N     = 20;    // Kugelanzahl
const auto MASSE = 0.01;  // Kugelmasse
const auto FEDER = 1.0;   // Federkonstante
const auto FREQ  = 0.5;   // Anregungsfrequenz
const auto AMPL  = 1.0;   // Anregungsamplitude
const auto REIB  = sqrt(FEDER * MASSE);  // Reibung

Matrix M(2 * N);
Vector b(2 * N);

// DGL-System der Wellenmaschine
Vector sys(double t, Vector u)
{
    double omega = 2 * M_PI * FREQ;
    return M * u + b * AMPL * sin(omega * t);
}

// Nach jedem Simulationsschritt: Kugelpositionen darstellen
void step(double t, Vector u)
{
    chart_line_series (u.Slice(0, N), CHART_BLACK);
    chart_point_series(u.Slice(0, N), CHART_RED  );
    chart_end_of_frame();
}

int main()
{
    // Matrix M belegen
    for(int i = 0; i < N; ++i)
    {
        M(i, N + i) = 1;
        M(N + i, i) = -2 * FEDER / MASSE;
        if(i > 0)     M(N + i, i - 1) = FEDER / MASSE;
        if(i < N - 1) M(N + i, i + 1) = FEDER / MASSE;
    }
    M(2 * N - 1, N - 1) = -FEDER / MASSE;
    M(2 * N - 1, 2 * N - 1) = -REIB / MASSE;
\end{lstlisting}
\pagebreak    
\MyListing{C++}
    // Vektor b belegen
    b(N) = FEDER / MASSE;

    // DGL-System integrieren
    Vector init(2 * N);
    dgl_rk4(0, 100, 2000, sys, step, init);

    // Animation der Simulationsergebnisse
    chart_frame_timer(100);
    chart_show(CHART_NOBACKGROUND, "Wellenmaschine");
}
\end{lstlisting}

\begin{figure}[t] 
  \centering 
  \includegraphics[width=1.0\linewidth]{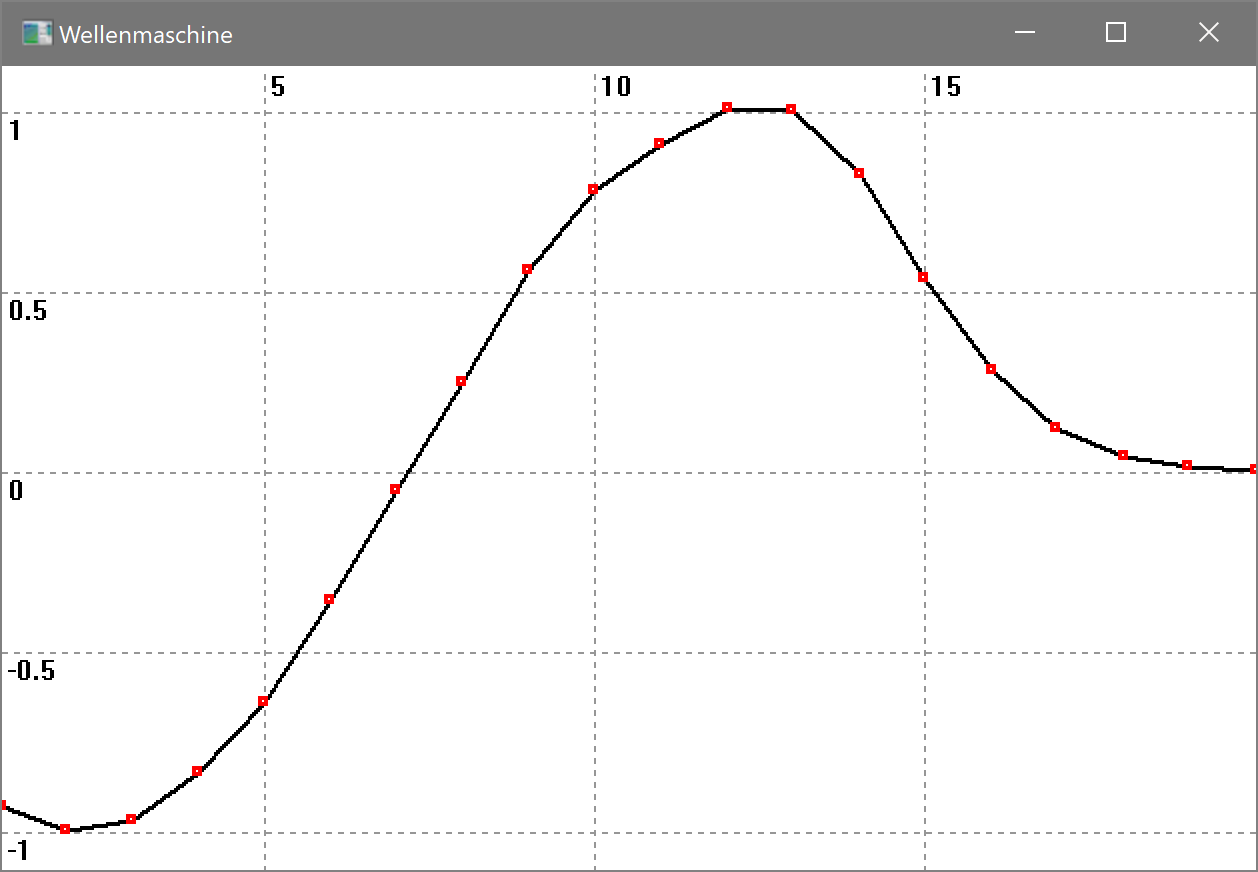}
  \caption{Simulation mit C++}
\end{figure} 

\section{Schlussbemerkungen}

Lineare Algebra, numerische Integration von Differentialgleichungen,
grafische Darstellung von Simulationsergebnissen -- all dies sind
Standardaufgaben in der Ingenieurinformatik. Die dazu notwendigen
Werkzeuge, entsprechende Bibliotheken usw. sind für alle in Technik
und Wissenschaft relevanten Programmiersprachen verfügbar. Für weit
verbreitete Sprachen wie C++ existiert eine besonders große Auswahl
an solchen Werkzeugen -- oft in professioneller Qualität, zum Teil
zugeschnitten auf spezielle Anwendungen. Dies kann insbesondere
Einsteigern Schwierigkeiten bereiten, zum Beispiel, wenn Bibliotheken
zunächst auf konkrete Programmierumgebungen angepasst oder gesondert
installiert werden müssen. Hinzu kommt, dass manche Bibliotheken
nur im Rahmen ganz bestimmter Software- oder Hardwareumgebungen
eingesetzt werden können.

Auf der Homepage des Autors stehen freie Bibliotheken für
technisch-wis\-sen\-schaft\-liche Anwendungen in C/C++ zum
Download bereit, bei deren Entwicklung auf eine einfache
Anwendbarkeit (gerade auch) durch Einsteiger geachtet wurde
(\href{http://kuepper.userweb.mwn.de/software.htm}%
{http://kuepper.userweb.mwn.de/software.htm}).
Bis auf wenige Ausnahmen sind diese Bibliotheken mit allen
relevanten PC-Be\-triebs\-syste\-men, Microsoft Windows, Mac OS X und
GNU/Linux kompatibel. Sie werden für das Programmbeispiel
in Abschnitt 4 genutzt.

\subsection*{Ausführungsgeschwindigkeit im Vergleich}

Neben den gegebenenfalls anfallenden Kosten für Softwarelizenzen
kann die Ausführungsgeschwindigkeit der fertig entwickelten Programme
ein weiterer wichtiger Aspekt bei der Auswahl einer Programmiersprache
bzw. -umgebung sein. Der Voll\-stän\-digkeit halber sei daher zu den
abgedruckten Programmbeispielen der jeweils gemessene Zeitbedarf für die
numerische Integration des Differentialgleichungssystems angegeben,
also die Dauer des ode45- (MATLAB/Octave), des integrate.odeint-
(Python) bzw. des dgl\_rk4-Aufrufs (C++). Die Messungen wurden auf
einem HP Spectre 13 mit i7-4500U-Prozessor, \SI{8}{GB} RAM und 
dem Betriebssystem Microsoft Windows 10 durchgeführt. Die Tabelle
zeigt jeweils die Mittelwerte aus 10 Messungen.

\begin{center}
\begin{tabular}{@{}l|r}
  MATLAB R2017a (64-bit)                          &
  \SI{641}{ms}                                    \\ \hline
  
  GNU Octave 4.0.0                                &
  \SI{4988}{ms}                                   \\ \hline

  Python 3.5.2 (64-bit)                           &
  \SI{346}{ms}                                    \\ \hline

  C++ (Visual Studio 2017, Release-Modus)         &
  \SI{16}{ms}
\end{tabular}
\end{center}

\newpage


\end{document}